\documentclass[lettersize,journal]{IEEEtran}
\usepackage[utf8]{inputenc}
\usepackage[T1]{fontenc}
\usepackage{amsmath,amsfonts}

\usepackage{algorithm}
\usepackage{algpseudocode}
\usepackage{tikz}
\usepackage{array}
\usepackage[caption=false,font=small,labelfont=rm,textfont=rm]{subfig}
\usepackage{textcomp}
\usepackage{stfloats}
\usepackage{url}
\usepackage{verbatim}
\usepackage{graphicx}
\usepackage{cite}
\usepackage{multirow}
\usepackage{booktabs}
\usepackage{makecell}
\usepackage{amssymb} 
\usepackage{cases} 
\usepackage{hyperref}  
\usepackage{tabularx}
\hypersetup{
	colorlinks=true,
	linkcolor=blue,
	anchorcolor=blue,
	citecolor=blue}
\usepackage{booktabs}
\usepackage[table]{xcolor}
\usepackage{pifont}
\usepackage{float}
\usepackage{threeparttable}

\begin{document}
	
	\title{ RF Instrument Agent (RFIA): Empowering RF Instruments with Natural Language  Understanding, Scheduling and Execution of Complex Tasks}
	
	\author{Chunhui Li, \textit{Graduate Student Member, IEEE}, and Wei Fan, \textit{Senior Member, IEEE}

		\thanks{This work was supported by the National Natural Science Foundation of
			China under Grant 62571119. \textit{(Corresponding author: Wei Fan.)}}
		\thanks{Chunhui Li and Wei Fan are with the National Mobile Communications Research Laboratory, School of Information Science and Engineering, Southeast University, Nanjing 210096, China, and also with the Purple Mountain Laboratories, Nanjing 211111, China (e-mail: lichunhui@seu.edu.cn; weifan@seu.edu.cn).} 
	}
	

	
	
	\maketitle

\begin{abstract}
Modern radio-frequency (RF) instruments, such as vector network analyzers (VNAs), already provide mature remote-control interfaces. However, practical RF measurement workflows still rely on manual operation or custom scripting, which is time-consuming and expertise-intensive. This paper presents RF Instrument Agent (RFIA), a natural-language agent framework for reliable task-driven RF instrument control. RFIA adopts a decoupled intent--planning--execution architecture, where the LLM is used only for task understanding and high-level planning, while instrument-facing operations are handled by a deterministic runtime. Verified skills, workflow templates, RF analysis tools, instrument-specific rules, and retrieval-assisted SCPI knowledge are organized in a structured knowledge base, and hybrid execution graphs are used for closed-loop measurement tasks. A hardware-in-the-loop prototype is implemented on a commercial VNA and evaluated using a 16-task benchmark covering configuration, query, acquisition, rule-aware operation, RF-data analysis, and closed-loop measurement. RFIA handles all benchmark tasks under predefined execution and safety policies, including one expected safety rejection. Hardware-in-the-loop results with both a 230B-scale MiniMax-M2.7 model and a smaller 27B-scale Qwen3.6-27B model confirm that the decoupled architecture supports reliable natural-language RF measurement automation across different LLM backends.
\end{abstract}

	\begin{IEEEkeywords}
		Radio-frequency (RF) instrumentation, intelligent measurement, automated test, agentic control
	\end{IEEEkeywords}

	\section{Introduction}
Radio-frequency (RF) instruments are a cornerstone of modern RF testing and measurement, and have long served as essential enabling tools for the evolution of wireless communication systems\cite{vidotto2022software}. In particular, instruments such as vector network analyzers (VNAs), spectrum analyzers, and signal generators provide the measurement foundation for the design, calibration, verification, and maintenance of RF devices and systems\cite{del2025virtual}. Despite the maturity of modern instrument hardware and remote-control interfaces, RF instrument operation still relies heavily on either manual interaction or automated scripts written with standard commands for programmable instruments (SCPI) \cite{mueller1989efficient, scpi1999standard}. Although script-based automation improves repeatability, it remains labor-intensive and often requires substantial expertise in instrument control, measurement procedures, and task-specific workflow design \cite{xie2025toward}. As a result, efficient execution of complex RF measurement tasks continues to depend strongly on experienced human operators.

The idea of using natural language to interact with measurement systems has a long history in the instrumentation community, with early work exploring learning-capable natural-language interfaces for computer-based measurement systems \cite{mangiavacchi1990innovative}.
Recent advances in large language models (LLMs) and artificial intelligence (AI) agent systems have created new opportunities for intelligent RF instrumentation. LLMs provide natural-language understanding, semantic abstraction, and high-level reasoning capabilities, while agent systems further enable task decomposition, tool orchestration, and execution management \cite{wang2024survey,wei2022chain}.  Recent studies have also explored LLM-assisted control-code generation and agentic operation of scientific instruments \cite{febba2025text,vriza2026operating}, indicating a broader shift from command-level scripting toward task-level automation.
A similar trend has appeared in the test-and-measurement industry, where representative examples include AI assistants for LabVIEW/TestStand \cite{niNigel}, AI-based design assistants for Keysight ADS \cite{keysightEDAAI}, AI-assisted scripting for Rohde \& Schwarz CMX500-based testing \cite{rohdeCMX500AI}, and Moku AI or generative instrumentation from Liquid Instruments \cite{goldberg2024please,liquidGenerativeInstrumentation}. Despite this progress, existing efforts mainly focus on code generation, scripting assistance, design support, or conversational guidance, and do not explicitly provide an execution-safe agent runtime for state-aware, rule-constrained, and closed-loop RF measurement. Therefore, the key gap is not the absence of remote-control interfaces, but the lack of a framework that bridges user intent and verified instrument operation.

Several fundamental challenges must be addressed to realize such an execution-safe RF instrumentation agent, as summarized below.

 First, there exists a substantial reliability gap between natural-language task understanding and deterministic instrument execution. User instructions are often open-ended, underspecified, and task-oriented, whereas RF instrument control requires strict command semantics, state-dependent execution, and deterministic behavior \cite{niu2026optics}. Since LLM generation is probabilistic, directly translating natural-language instructions into SCPI commands or control code can easily lead to execution failures, invalid state transitions, and accumulated errors.

Second, existing general-purpose agent frameworks and multi-agent orchestration paradigms, such as ReAct-style tool-use agents, are not specifically designed for instrument-oriented execution~\cite{yao2022react}. Although such frameworks support task decomposition and external tool invocation, they typically lack the structured knowledge organization, explicit state management, execution constraints, and conflict-aware control logic required by RF instruments. In practice, RF instruments involve large command spaces, strong inter-command dependencies, and complex internal states. As a result, even when a general-purpose agent correctly interprets user intent, it may still fail to complete the task reliably because it cannot guarantee instrument-safe planning and verifiable execution.

Third, the knowledge organization problem for RF instrument agents remains largely unexplored. Modern RF instruments expose thousands of SCPI commands with heterogeneous functions, hierarchical structures, and nontrivial coupling relationships \cite{arpaia2011model}. For an agent, the key challenge is not only how to retrieve individual commands, but also how to organize, select, and compose them into task-relevant executable knowledge under limited context budgets \cite{lewis2020retrieval}. 

Fourth, many practical RF measurement tasks are not static command sequences, but dynamic closed-loop workflows \cite{wu2025adaptive}. Their execution often depends on intermediate measurement results and may require data acquisition, local analysis, conditional branching, parameter adjustment, and repeated measurements. Such workflows are difficult to handle with conventional scripts and cannot be adequately supported by naive tool-calling mechanisms alone.

 To address the above challenges, this paper proposes the RF Instrument Agent (RFIA), a dedicated agent framework for intelligent RF instrument control. Unlike generic agent frameworks, RFIA is specifically designed for instrument-oriented execution and remains fully compatible with existing SCPI-enabled instruments.
 The main contributions of this paper are summarized as follows.

\begin{itemize}
	\item We propose RFIA, a dedicated agent framework for task-driven RF instrument control. RFIA adopts a decoupled intent--planning--execution architecture that confines the LLM to task understanding and high-level planning, while delegating all instrument-facing operations to a deterministic and verifiable runtime. By preventing the LLM from directly generating code or SCPI commands, RFIA reduces token usage, supports state-aware and rule-aware orchestration, and is compatible with lightweight LLM backends.

	\item We design a structured instrument knowledge organization mechanism for RFIA. Verified skills, workflow templates, RF analysis tools, instrument-specific rules, and retrieval-assisted SCPI knowledge are organized as reusable knowledge units and progressively disclosed according to task requirements. This mechanism supports frequent operations through prevalidated skills, long-tail requests through controlled retrieval, and compact LLM interaction without direct code or SCPI generation.

	\item We implement a hardware-in-the-loop RFIA prototype on a commercial VNA and design a 16-intent benchmark covering configuration, query, routine acquisition, rule-aware operation, RF-data analysis, closed-loop measurement, and multi-segment data logging. The prototype is evaluated with both MiniMax-M2.7 and Qwen3.6-27B under the same RFIA runtime, and both hardware-in-the-loop runs complete all benchmark tasks under predefined execution and safety policies, including one expected safety rejection.
\end{itemize}

\begin{figure*}[!t]
	\centering
	\includegraphics[width=1\linewidth]{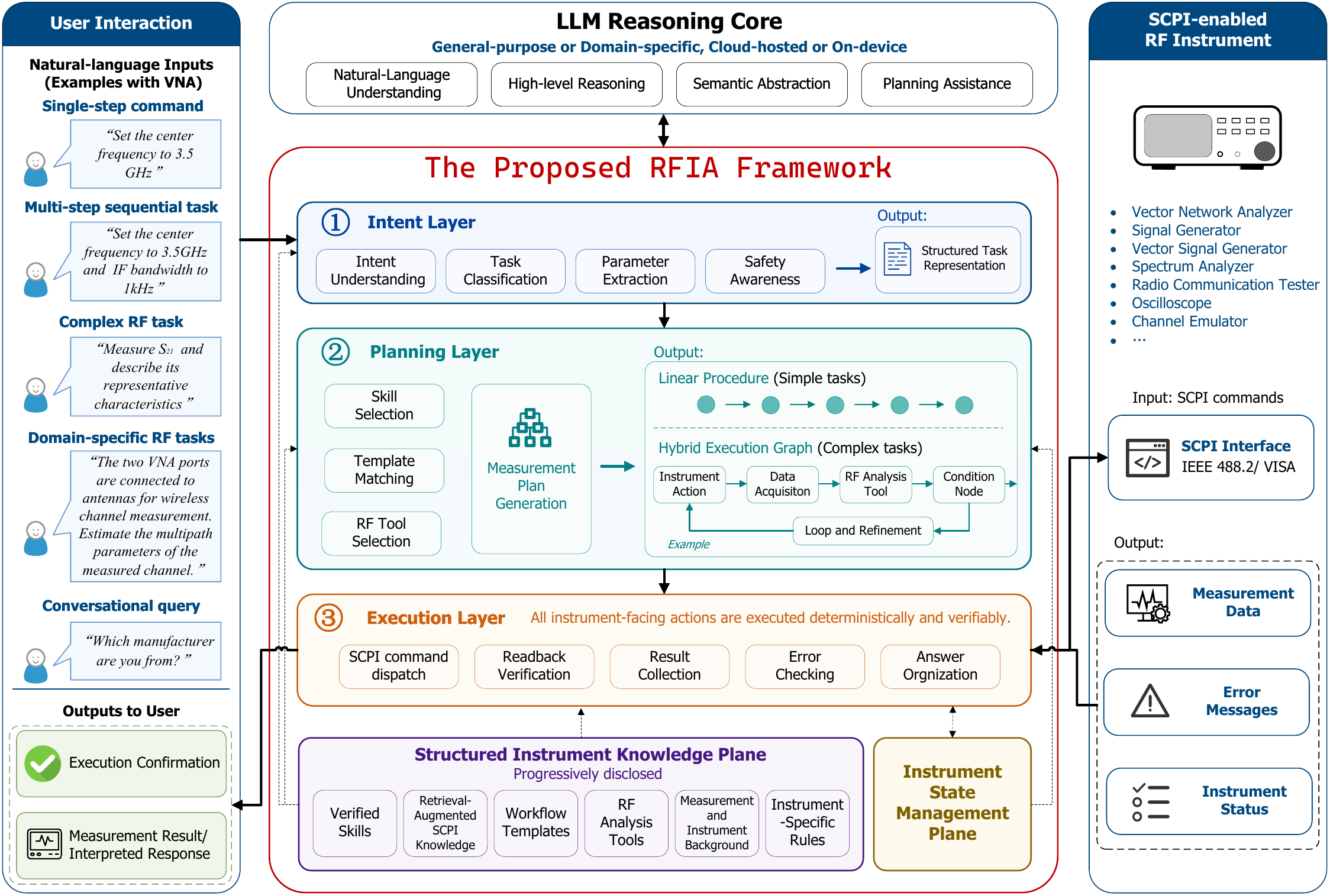}
	\caption{Overall architecture of the proposed RFIA framework for natural-language-driven RF instrument control.}
	\label{fig:rfia_architecture}
\end{figure*}

\section{Design Principles and Architecture Overview }

\subsection{Design Objectives}

The objective of RFIA is to enable reliable and efficient task-driven control of RF instruments from natural-language user instructions. More importantly, RFIA is not intended merely to directly map user input to instrument commands, but to bridge the gap between high-level task intent and low-level deterministic execution. In the following, a vector network analyzer (VNA) is taken as a representative example to illustrate the target capabilities of RFIA. Accordingly, RFIA is designed to satisfy the following objectives.

\begin{enumerate}
	\item \textbf{Single-step instruction execution:} RFIA should accurately interpret a single natural-language instruction and map it to the corresponding SCPI command for execution. For example, for a VNA, the instruction \textit{``Set the center frequency to 3.5~GHz''} should be correctly translated into the corresponding instrument operation.
	
	\item \textbf{Multi-step sequential task execution:} RFIA should support natural-language requests that correspond to an ordered sequence of multiple SCPI operations. For example, for a VNA, the instruction \textit{``Set the center frequency to 3.5~GHz and adjust the IF bandwidth to 1~kHz''} should be converted into a valid command sequence and executed in the correct order.
	
	\item \textbf{Complex task execution with RF analysis:} RFIA should support tasks that go beyond direct command arrangement and involve RF measurement acquisition, local signal analysis, and semantic interpretation of results. For example, for a VNA, RFIA should be able to execute the instruction \textit{``Measure \(S_{21}\) and describe its representative characteristics''} by performing the measurement, invoking the required analysis tools, and returning an interpretable result.
	
	\item \textbf{Domain-specific RF tasks:} RFIA should further support expert-level RF tasks that combine instrument control, measurement data acquisition, and domain-specific signal processing. For example, when two VNA ports are connected to antennas for wireless channel measurement, RFIA should be able to handle an instruction such as \textit{``The two VNA ports are connected to antennas for wireless channel measurement. Estimate the multipath parameters of the measured channel.''} This requires RFIA not only to configure and query the instrument, but also to invoke specialized RF processing algorithms to infer physically meaningful channel parameters from the acquired data.
	
	\item \textbf{Conversational interaction:} RFIA should also support general conversational queries related to the instrument, such as \textit{``Which manufacturer are you from?''}, thereby providing a more natural human--instrument interaction interface.
\end{enumerate}

In addition to functional coverage, RFIA is designed to satisfy several system-level requirements. It should achieve high execution accuracy and reliability while remaining lightweight and efficient in practical deployment. It should be extensible to newly introduced instrument commands and newly developed complex-task capabilities without requiring fundamental redesign of the framework. It should also be broadly applicable to a wide range of SCPI-enabled instruments, rather than being restricted to a single device model. Therefore, an important design objective of RFIA is to support scalable access to large SCPI command spaces while continuously improving its capability for complex task execution through structured knowledge expansion and runtime evolution.

\subsection{Overall Architecture of RFIA}

Fig.~\ref{fig:rfia_architecture} illustrates the overall architecture of the proposed RFIA framework. RFIA is an LLM-centered yet execution-decoupled agent framework for SCPI-enabled RF instruments, aiming to bridge natural-language task intent and deterministic instrument execution. Rather than directly translating user instructions into SCPI commands, RFIA introduces an intermediate task representation and organizes the control process into three functional layers and two supporting planes. The three functional layers are the intent layer, the planning layer, and the execution layer, while the two supporting planes are the structured instrument knowledge plane and the instrument state management plane.

The intent layer interprets the user instruction and converts it into a structured task representation, where the task type, target quantity, key measurement parameters, safety constraints, and feedback requirements are explicitly specified. Based on this representation, the planning layer generates an executable task path according to the task type, current instrument state, available skills, and required analysis tools. For simple configuration, query, or routine measurement tasks, the task path is represented as a linear procedure composed of ordered executable steps. For complex closed-loop measurement tasks, the task path is formulated by instantiating a registered hybrid execution-graph template that integrates instrument-control, data-acquisition, RF-analysis-tool, condition, and loop-refinement nodes. The execution layer then performs instrument-facing actions in a deterministic and verifiable manner, including SCPI command dispatch, readback verification, error checking, state updating, and result collection.

The two supporting planes provide the knowledge and runtime context required by the three-layer workflow. The structured instrument knowledge plane organizes verified skills, retrieval-augmented SCPI knowledge, workflow templates, RF analysis tools, measurement background, and instrument-specific rules into reusable knowledge units. This design allows high-frequency operations to be handled through lightweight prevalidated skills, while long-tail requests are supported through retrieval-augmented access to the broader SCPI command space. The instrument state management plane maintains the current configuration, measurement state, execution context, and historical observations of the instrument. These layers and planes are detailed in the following sections.

\section{Three-Layer Functional Architecture}
Following the overall architecture in Fig.~\ref{fig:rfia_architecture}, this section further specifies the responsibility boundaries and interface contracts among the three functional layers. The key design principle is that the output of each layer should be sufficiently structured for the next layer to inspect, constrain, and verify, rather than directly triggering lower-level instrument actions. Specifically, the intent layer produces a structured semantic contract rather than SCPI commands; the planning layer produces an executable task structure rather than instrument I/O; and the execution layer is the only component allowed to interact with the physical instrument through a deterministic and verifiable runtime. This layered contract prevents unconstrained LLM outputs from directly reaching the instrument, while still allowing semantic reasoning and task planning to guide complex RF measurement workflows.

\subsection{Intent Layer}

The intent layer is responsible for grounding a natural-language user request into a structured task contract for the planning layer. Specifically, it performs three functions: task classification, parameter grounding, and safety-aware constraint marking.

Task classification is not used merely as a semantic label, but as a routing decision for the agent runtime. The classified task type determines the downstream planning route, the required knowledge resources, the executable structure to be generated, and the expected output format. To avoid unconstrained task interpretation, the classification process is guided by the structured instrument knowledge plane, which provides task definitions, skill metadata, workflow templates, RF analysis tool descriptions, and safety rules. As summarized in Table~\ref{tab:intent_taxonomy}, RFIA adopts an execution-oriented task taxonomy defined according to the operational role of a user request in the agent workflow, rather than the command set of a particular instrument. Conversational tasks are handled by response generation without triggering instrument execution. State query and configuration tasks are routed to validated skills, with the latter requiring parameter validation, readback verification, and state updating. Acquisition tasks are handled by predefined measurement workflows, whereas analysis tasks require RF analysis tools and result formatting. Feedback-control tasks are routed to hybrid execution graphs because their subsequent actions depend on intermediate observations, analysis results, validation feedback, or iterative refinement.

\begin{table}[t]
	\centering
	\caption{Execution-oriented task taxonomy used in the intent layer.}
	\label{tab:intent_taxonomy}
	\footnotesize
	\renewcommand{\arraystretch}{1.18}
	\setlength{\tabcolsep}{3pt}
	\begin{tabularx}{\columnwidth}{p{2.1cm} p{2.3cm} X}
		\toprule
		\textbf{Task Class} 
		& \textbf{Planning Route} 
		& \textbf{Expected Output} \\
		\midrule
		
		Conversational task
		& Response generation
		& Natural-language response based on instrument profile or background knowledge. \\
		
		State query task
		& Direct query skill
		& Current instrument state, setting, or response. \\
		
		Configuration task
		& Direct skill or linear workflow
		& Verified state update after validation and readback. \\
		
		Acquisition task
		& Linear acquisition workflow
		& Measurement trace, sampled data, or stored file. \\
		
		Analysis task
		& Tool-augmented workflow
		& Physical or statistical interpretation of acquired data. \\
		
		Feedback-control task
		& Hybrid execution graph
		& Refined setting, optimized condition, or target measurement result. \\
		
		\bottomrule
	\end{tabularx}
\end{table}

Parameter grounding extracts instrument-relevant slots from the user request and converts them into canonical fields for the planning layer, without generating low-level instrument commands. Typical fields include frequency range, source power, sweep points, port configuration, measurement quantity, and output format. Missing or ambiguous fields are explicitly marked. Safety-aware constraint marking flags potentially risky operations, such as increasing source power, enabling RF output, performing wideband sweeps, or overwriting calibration/data files, so that they can be checked by the planning and execution layers before physical operation.

As an example, consider the instruction to a signal generator: 
``\textit{Set the carrier frequency to 2.6~GHz, increase the output power to 10~dBm, and turn on RF output.}''
The intent layer classifies this request as a configuration task and grounds the key parameters into canonical fields, including carrier frequency, output power, and RF output state. Since the request involves power modification and RF-output enabling, safety flags for power-limit checking, RF-output checking, and readback verification are attached to the task contract.

In implementation, the intent layer is realized through schema-constrained prompting. The LLM is instructed to generate only a JSON-formatted task contract that follows a predefined schema. It is explicitly prohibited from generating raw SCPI commands, executable code, or direct instrument I/O operations. The prompt is supplied with structured metadata from the instrument knowledge plane, including task definitions, available skills, workflow templates, RF analysis tools, and safety rules. In this way, task classification is performed as a constrained routing decision rather than an unconstrained semantic judgment.

The resulting task contract is constrained by both the predefined output schema and the available structured knowledge units. This improves the consistency of task routing, prevents unsupported or unsafe operations from being passed to the planning layer, and enables downstream resources to be selected only from verified skills, workflow templates, and RF analysis tools.

\subsection{Planning Layer}

\begin{figure}[t]
	\centering
	\includegraphics[width=\columnwidth]{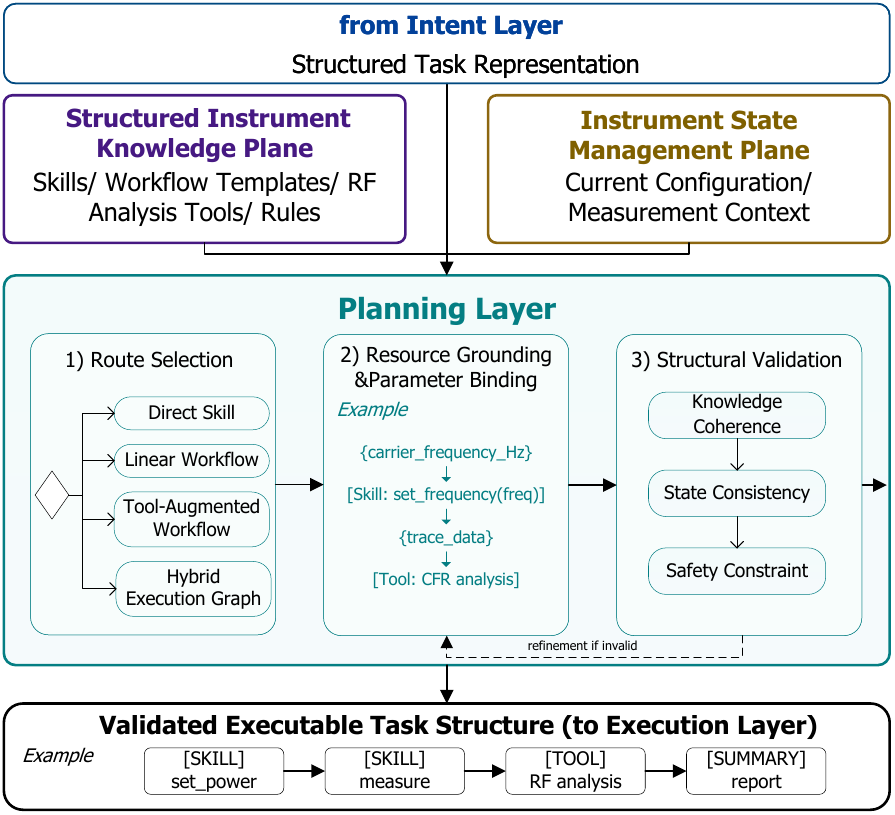}
	\caption{Planning layer as a constrained compilation bridge. }
	\label{fig:planning_layer_compilation}
\end{figure}

The planning layer functions as a reliability bridge between the structured intent contract and the deterministic execution runtime. Its responsibility is to transform the task contract generated by the intent layer into a validated executable task structure, without generating executable code, issuing SCPI commands, or communicating with the physical instrument. To achieve this, the planning layer uses three sources of information: the structured task contract from the intent layer, the structured instrument knowledge plane, and the instrument state management plane. The knowledge plane provides task definitions, validated skills, workflow templates, RF analysis tools, and instrument-specific rules, while the state management plane provides the current configuration, active measurement context, and state-dependent constraints. In this sense, the planning layer acts as a constrained compilation layer: it reasons about task organization and resource selection under knowledge and state constraints, while leaving all instrument-facing operations to the execution layer. 
As illustrated in Fig.~\ref{fig:planning_layer_compilation}, the compilation process consists of three stages: planning route selection, resource grounding and parameter binding, and structural validation.

\textbf{1) Planning Route Selection:}
Given a task contract, the planning layer first selects an appropriate planning route according to the task class, expected output, safety constraints, available resources, and current instrument state. This selection is guided by the task taxonomy and resource metadata maintained in the structured instrument knowledge plane, rather than by unconstrained LLM judgment. The route determines the complexity and topology of the executable task structure:
\begin{itemize}
	\item \textbf{Direct Skill Route:} State-query tasks and simple configuration tasks are routed to a single validated skill. This route provides low latency and high reliability for high-frequency operations whose execution logic is fixed in advance.
	
	\item \textbf{Linear Workflow Route:} Multi-parameter configuration tasks and acquisition tasks are routed to ordered workflow templates composed of validated skill nodes. This route is used when several instrument operations must be performed in a predefined sequence with explicit inter-step dependencies and verification requirements.
	
	\item \textbf{Tool-Augmented Workflow Route:} Analysis tasks are routed to workflows that combine data acquisition or data loading with local RF analysis tools and result-formatting modules. This route allows measurement data to be converted into physically meaningful results without embedding analysis logic into instrument-control commands.
	
	\item \textbf{Hybrid Execution Graph Route:} Feedback-control tasks are routed to registered hybrid execution-graph templates. Such templates may contain instrument-control nodes, acquisition nodes, RF-analysis nodes, condition nodes, and refinement nodes. This route is used when subsequent actions depend on intermediate observations, analysis results, validation feedback, or iterative refinement.
\end{itemize}
By separating these routes, RFIA uses the simplest executable structure that satisfies the task requirement, while reserving graph-based planning for tasks that require feedback-dependent reasoning.

\textbf{2) Resource Grounding and Parameter Binding:}
After route selection, the planner grounds the selected route into concrete resources retrieved from the structured instrument knowledge plane and binds the grounded parameters in the task contract to the input slots of validated skills, workflow templates, and analysis tools. This step selects the applicable resource instances for the target instrument, resolves symbolic references, and checks parameter completeness before execution. For example, a grounded field such as \texttt{carrier\_frequency\_Hz} can be bound to the frequency input of a signal-generator configuration skill, while a measured trace can be bound to the input of an RF analysis tool. The output of this stage is a self-contained task structure in which each node has explicit inputs, expected outputs, and required verification conditions.

\textbf{3) Structural Validation:}
Before dispatching the task structure to the execution layer, the planning layer validates it against the structured instrument knowledge plane, the instrument state management plane, and the safety constraints inherited from the intent layer. The knowledge plane confirms that the referenced skills, workflow templates, and analysis tools are available for the target instrument and applicable to the selected route. The state manager checks whether the selected resources are compatible with the current configuration and active measurement context, and identifies state-dependent prerequisites or potential conflicts. Safety constraints, such as power-limit checking, RF-output checking, calibration protection, and file-overwrite protection, are propagated into mandatory validation nodes or runtime checks. In this way, safety-related intent is not only recorded as metadata, but also embedded into the executable structure that will be verified during execution.

The final output of the planning layer is a resolved and pre-validated executable task structure. It specifies the resources to be used, the ordering of their invocation, the bound parameters, and the checks required before or during execution. This design allows RFIA to exploit LLM-based reasoning for task orchestration while keeping the physical instrument control path deterministic, constrained, and verifiable.

RFIA does not ask the LLM to synthesize arbitrary internal control flow for hybrid execution graphs. When a validated workflow template is available, RFIA gives it priority and only allows the LLM to select the route and bind user-intended parameters. Condition nodes and loop structures are therefore pre-built, parameterized resources in the knowledge plane, such as a coarse-to-fine resonance-search template with configurable stopping criteria. This design preserves limited planning flexibility while keeping closed-loop execution deterministic and auditable. Unsupported closed-loop patterns must first be registered or validated before execution.

\subsection{Execution Layer}

The execution layer is the only instrument-facing component of RFIA. It receives a validated executable task structure from the planning layer and realizes it through a deterministic runtime. Unlike the upper layers, it performs no free-form reasoning, accepts no LLM-generated SCPI commands, and synthesizes no executable code; it only executes validated resources according to the provided task structure. In this way, all physical instrument operations remain constrained, auditable, and reproducible.

The runtime follows the node topology specified in the task structure. In RFIA, the executable task structure is represented by typed nodes, including instrument-control nodes, acquisition nodes, tool nodes, verification nodes, and condition/refinement nodes. {Instrument-control nodes} invoke verified skills that internally encapsulate the required SCPI sequences and command-ordering logic. {Acquisition nodes} collect instrument outputs such as traces, readbacks, or measurement status. {Tool nodes} execute local RF analysis functions on acquired data. {Verification nodes} perform readback checking, error querying, or output validation. {Condition/refinement nodes} evaluate structured observations against defined criteria and determine whether the graph continues, terminates, or enters another refinement step. Execution therefore always operates on typed task nodes, never on raw language-model outputs. The core runtime protocol is summarized in Algorithm~\ref{alg:execution_verify_commit}.

\begin{algorithm}[t]
	\caption{Deterministic runtime execution with verify-then-commit}
	\label{alg:execution_verify_commit}
	\footnotesize
	\begin{algorithmic}[1]
		\Require Validated executable task structure \(\mathcal{G}\); maintained instrument state \(S\)
		\Ensure Verified execution record \(\mathcal{R}\) or structured failure report \(\mathcal{F}\)
		
		\State \(\mathcal{R} \leftarrow \emptyset\)
		\State \(v \leftarrow \textsc{NextNode}(\mathcal{G}, S, \mathcal{R})\)
		
		\While{\(v \neq \mathrm{null}\)}
		\If{\(\textsc{PreCheck}(v,S) = \mathrm{fail}\)}
		\State \(\mathcal{F} \leftarrow \textsc{FailureReport}(v,\mathrm{precondition\_fail})\)
		\State \Return \(\mathcal{F}\)
		\EndIf
		
		\State \(o \leftarrow \textsc{ExecuteNode}(v,S)\)
		\Comment{skill, acquisition, tool, verification, or condition node}
		\State \(m \leftarrow \textsc{Verify}(v,o,S)\)
		
		\If{\(m \neq \mathrm{ok}\)}
		\State \(\mathcal{F} \leftarrow \textsc{FailureReport}(v,m,o)\)
		\State \(S \leftarrow \textsc{AbortOrMarkInvalid}(S,v)\)
		\State \Return \(\mathcal{F}\)
		\EndIf
		
		\State \(S \leftarrow \textsc{CommitState}(S,v,o)\)
		\State \(\mathcal{R} \leftarrow \mathcal{R} \cup \{(v,o,m)\}\)
		\State \(v \leftarrow \textsc{NextNode}(\mathcal{G},S,\mathcal{R})\)
		\EndWhile
		
		\If{\(\textsc{RequireSummary}(\mathcal{G})\)}
		\State \(\mathcal{R}.\mathrm{summary} \leftarrow \textsc{LLMSummarize}(\mathcal{R})\)
		\Comment{LLM invoked only on structured records}
		\EndIf
		
		\State \Return \(\mathcal{R}\)
	\end{algorithmic}
\end{algorithm}

A central principle of the execution layer is that command dispatch is not equivalent to successful execution. Each instrument-facing operation is followed by explicit verification, such as readback checking, instrument-error querying, or output-format validation. Once verified, the result is used to synchronize the instrument state management plane. If a verification mismatch, timeout, instrument error, or safety violation occurs, execution stops and a structured failure report is generated, specifying the failed node, error type, and recommended recovery action. This prevents unverified failures from being silently propagated through the system.

The execution layer also serves as the feedback interface for closed-loop tasks. Verified instrument readbacks, acquired traces, tool outputs, and condition results are captured as structured observations. These observations update the state management plane, drive conditional branching in hybrid execution graphs, and, when a user-facing explanation is needed, are supplied to the LLM for summarization---strictly after deterministic execution and validation. The LLM's role remains limited to interpreting structured observations and assisting high-level refinement; it is never granted access to the instrument control path.

As an illustration, in a tool-augmented measurement task, the runtime may acquire an \(S_{21}\) trace, pass it to a local analysis tool, validate the output, update the state management plane, and only then provide the structured result to the LLM for natural-language summarization. In a feedback-control task, acquisition and analysis nodes may be executed repeatedly until a condition node signals completion.

The output of the execution layer is therefore a verified execution record: confirmed state updates, data references, tool results, validation outcomes, and, when required, user-facing summaries. This closes the loop between the executable task structure and physical instrument behavior while preserving the essential architectural separation---LLM-based reasoning for task orchestration, deterministic and verifiable execution for instrument-facing operations.

\section{Structured Instrument Knowledge Plane}

The structured instrument knowledge plane supplies RFIA with typed, reusable, and verifiable resources that support all three layers.  Modern RF instruments expose large SCPI command sets, but practical usage is highly skewed: a small subset of high-frequency operations accounts for most routine work, while the remaining long-tail commands are rarely needed but must remain accessible.  RFIA addresses this asymmetry by separating high-frequency verified execution from retrieval-augmented long-tail access, and by extending the knowledge plane with workflow templates, RF analysis tools, and instrument-specific rules.

\subsection{Knowledge Resource Types and Their Roles}

The knowledge plane is organized around five resource types, as summarized in Table~\ref{tab:knowledge_resources}.
\begin{enumerate}
	\item \textbf{Verified skills} are the basic executable abstractions. Each skill encapsulates a validated instrument operation with a structured interface (semantic name, input schema, parameter constraints, preconditions, SCPI command sequence, verification rule, error-checking policy, state update rule, and safety tags). The intent and planning layers operate on skill metadata, while the execution layer invokes the skill through the deterministic runtime.  This separation prevents the LLM from ever generating low-level commands for frequent operations.
	\item \textbf{Workflow templates} encode reusable multi-step procedures as ordered skill nodes with explicit inter-step dependencies, default parameters, validation points, and failure-handling rules.  They avoid repeatedly assembling operation sequences and preserve known measurement structures.
	\item \textbf{RF analysis tools} are deterministic local functions that process measurement data into physically meaningful results (e.g., resonance search, passband estimation, multipath-parameter extraction).  They are registered with strict input/output schemas and can be inserted into tool-augmented workflows or hybrid execution graphs, extending RFIA's domain capability without embedding analysis inside the LLM.
	\item \textbf{Instrument-specific rules} capture valid parameter ranges, state-dependent prerequisites, command-ordering constraints, calibration and file-overwrite protections, safe power limits, and readback requirements.  These rules guide safety-aware marking in the intent layer, structural validation in the planning layer, and runtime checks in the execution layer.
	\item \textbf{SCPI documents} serve as retrieval-augmented reference knowledge for long-tail requests not yet covered by verified skills or templates.  Retrieved fragments inform task interpretation and planning, but any resulting operation must still pass through resource validation, parameter checking, safety constraints, and deterministic execution.
\end{enumerate}

\begin{table}[t]
	\centering
	\caption{Structured knowledge resources in RFIA.}
	\label{tab:knowledge_resources}
	\footnotesize
	\renewcommand{\arraystretch}{1.18}
	\setlength{\tabcolsep}{3pt}
	\begin{tabularx}{\columnwidth}{p{2.15cm} p{2.35cm} X}
		\toprule
		\textbf{Resource Type} 
		& \textbf{Role} 
		& \textbf{Typical Content} \\
		\midrule
		Verified skill
		& Stable execution interface for frequent operations
		& Input schema, command sequence, preconditions, verification rule, state update, safety tags. \\
		Workflow template
		& Reusable structure for multi-step tasks
		& Ordered skill nodes, dependencies, default parameters, validation points, failure-handling policy. \\
		RF analysis tool
		& Deterministic local processing for RF data
		& Input/output schema, data type, processing function, physical interpretation. \\
		Instrument-specific rule
		& Constraint source for safe and valid operation
		& Parameter ranges, state dependencies, ordering constraints, calibration protection, readback requirements. \\
		SCPI document
		& Long-tail reference knowledge
		& Vendor command reference, programming guide, model-specific notes, measurement background. \\
		\bottomrule
	\end{tabularx}
\end{table}

\subsection{Balancing Stability and Extensibility via Progressive Disclosure}

The knowledge plane is not exposed monolithically.  Instead, it adopts a progressive disclosure mechanism, as summarized as follows:
\begin{itemize}
	\item During \textbf{intent grounding}, only task definitions, task classes, route policies, and summaries of safety rules are shown, constraining classification and safety marking.
	\item During \textbf{planning}, candidate skill, template, and tool metadata are disclosed so that the planner can select resources, bind parameters, and build executable structures under instrument-state constraints.
	\item During \textbf{execution}, the full executable specifications of the selected resources are used by the deterministic runtime, but these low-level details are never exposed as free-form generation targets for the LLM.
	\item \textbf{Long-tail retrieval} is invoked on demand and only the relevant SCPI reference fragments are provided to aid interpretation, while all safety and validation mechanisms remain enforced.
\end{itemize}
This staged disclosure improves context efficiency and prevents irrelevant knowledge from interfering with LLM reasoning, while ensuring that every operation is ultimately grounded in registered, verifiable resources.

All resources are registered through a standardized interface (name, functional description, supported instrument, input/output schemas, preconditions, dependencies, validation rules, safety tags, and execution type).  New capabilities—whether adding an instrument command, a measurement procedure, an analysis algorithm, or a safety policy—can be introduced by registering new resources without redesigning the agent framework.  In summary, the structured instrument knowledge plane acts as a typed, progressively disclosed resource space that separates stable execution from long-tail retrieval, anchors LLM reasoning to verifiable operations, and enables domain extension through registered skills, templates, tools, and rules.

\subsection{Instrument State Management Plane}

The instrument state management plane maintains the runtime context required for safe and state-consistent instrument control.  It records the confirmed instrument configuration (frequency, power, sweep parameters, active ports, measurement mode, output state), the execution context (active task or graph node, acquired data references, tool outputs, validation results), and safety-related information (protected calibration states, file-overwrite status, safe operating limits, unresolved failure conditions).  All state fields are updated only after successful deterministic verification in the execution layer, following a verify-then-commit discipline.

This explicit state object supports all three functional layers.  In the intent layer, it helps identify missing or ambiguous user requests by providing the current configuration as context.  In the planning layer, it is queried to check whether selected skills, workflow templates, and RF tools are compatible with the current instrument state.  In the execution layer, it is synchronized only after readback confirmation, error-free execution, and validation success.  If a verification mismatch, timeout, instrument error, or safety violation occurs, the affected state is marked as invalid rather than silently propagated, enabling recovery, replanning, or closed-loop refinement.  Together with the structured knowledge plane, the state management plane provides the runtime memory that keeps RFIA's operation grounded in verified physical instrument status rather than LLM-inferred assumptions.

\section{Experimental Validation}

\subsection{Prototype Implementation}

\begin{figure}[t]
	\centering
	\includegraphics[width=0.92\columnwidth]{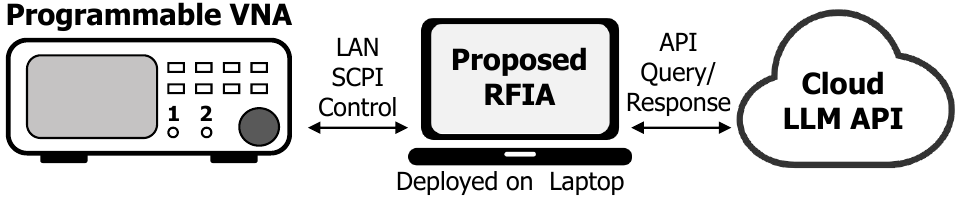}
	\caption{Hardware-in-the-loop architecture of the RFIA prototype. The commercial VNA is controlled by the local RFIA runtime through a LAN-based SCPI interface, while the cloud-hosted LLM is accessed only through the agent interface and has no direct access to instrument I/O.}
	\label{fig:vna_prototype}
\end{figure}

A hardware-in-the-loop prototype is implemented to validate the proposed RFIA framework. A commercial Ceyear 3671C VNA is used as the representative SCPI-enabled RF instrument, and an edge laptop is used as the local host for agent execution. As shown in Fig.~\ref{fig:vna_prototype}, the VNA is connected to the laptop through a LAN interface, while the LLM is accessed through a cloud-hosted API. The hardware experiments are conducted with MiniMax-M2.7 and repeated with the SiliconFlow-hosted Qwen3.6-27B model under the same RFIA runtime, in order to evaluate whether the verified execution architecture remains effective with a smaller model. The LLM is isolated from the VNA communication interface, and all instrument-facing operations are performed by the local deterministic runtime. For each benchmark intent, the system records the raw LLM intent output, the normalized executable intent, the generated skill/tool nodes, token usage, agent planning latency, local execution latency, and final runtime artifacts. The physical measurement results reported in the case studies are obtained from the corresponding measurement setups, while software-overhead statistics are extracted from the recorded RFIA project logs.

\begin{figure}[t]
	\centering
	\includegraphics[width=\columnwidth]{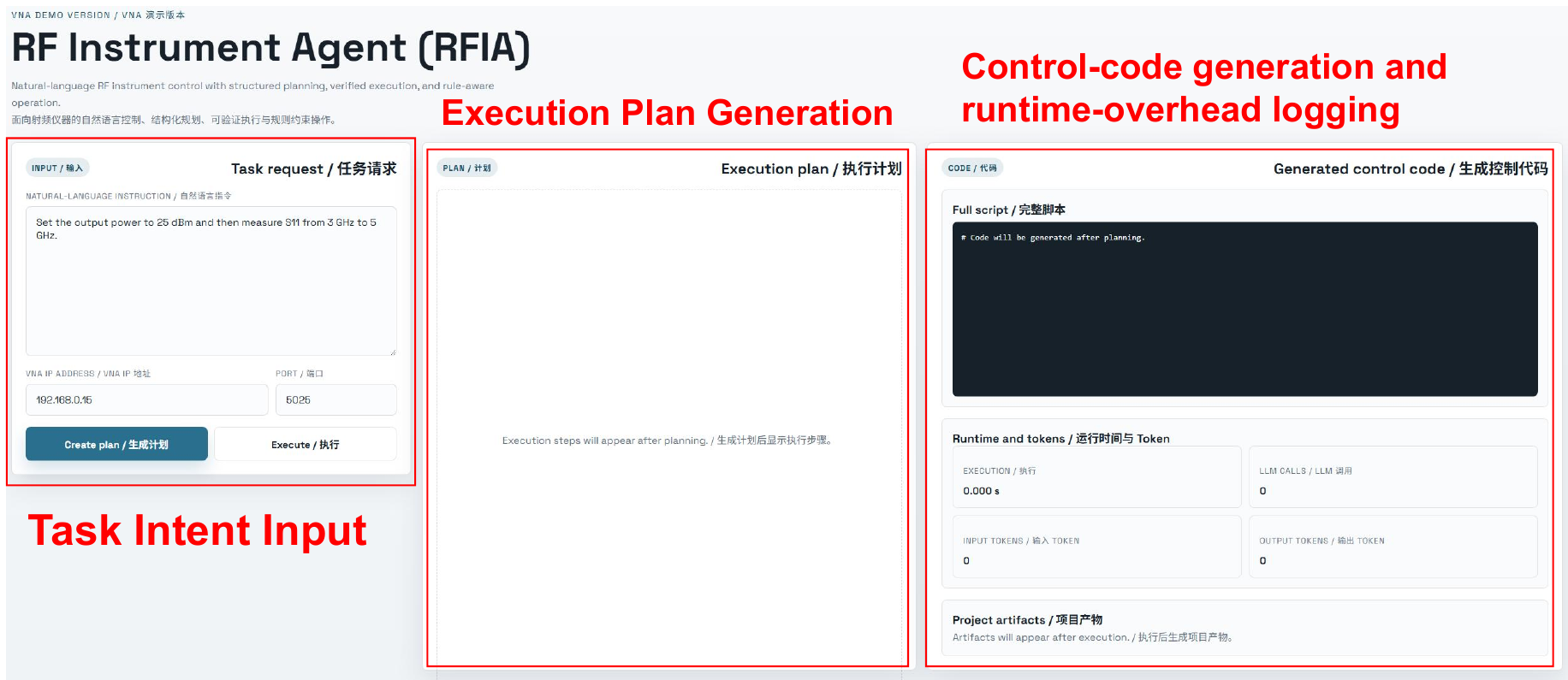}
	\caption{Implementation frontend of the RFIA, consisting of three modules: Task intent input, execution plan generation, and control code generation with runtime overhead logging.}
	\label{fig:frontend}
\end{figure}

The RFIA runtime deployed on the edge laptop includes the intent layer, planning layer, execution layer, structured instrument knowledge plane, and instrument state management plane. Instrument control is implemented through Python-based SCPI communication over LAN. High-frequency VNA operations, such as center-frequency configuration, span setting, sweep-point configuration, IF bandwidth setting, output-power query, trace acquisition, and \(S\)-parameter data reading, are encapsulated as verified skills. Each skill provides a structured input interface and internally contains the corresponding SCPI command sequence, readback query, error checking, and state update rule. Therefore, common instrument operations are executed through stable and prevalidated interfaces rather than through LLM-generated commands. The frontend of the RFIA is illustrated in Fig. \ref{fig:frontend}.

To support long-tail instrument functions beyond the predefined skill set, the VNA programming manual is organized into a retrieval-augmented SCPI command knowledge base. The manual content is parsed, normalized, and categorized into structured command entries, which are indexed in a vector database for semantic retrieval during runtime. This knowledge base provides reference information for intent grounding and planning, but retrieved commands are not directly executed by the LLM. Any operation inferred from the retrieved command knowledge must still be grounded into the RFIA task structure and executed through the deterministic runtime with parameter checking, readback verification, and state synchronization.

The prototype also includes workflow templates and local RF analysis tools. Workflow templates are used for multi-step measurement procedures, such as \(S_{11}\) or \(S_{21}\) acquisition over a specified frequency range. Local RF analysis tools support trace feature extraction, magnitude-range calculation, peak or minimum detection, time-domain transformation, and multipath-parameter estimation. These resources allow RFIA to support not only basic instrument configuration, but also measurement interpretation and feedback-driven RF tasks.

During evaluation, the prototype records an execution log for each user intent. The log includes the original natural-language instruction, structured task contract, selected planning route, invoked skills or workflow templates, SCPI-level execution status, readback values, instrument-error responses, acquired data references, local tool outputs, structured observations, and final user-facing response. These logs are used in the following subsections to evaluate task success, routing correctness, safety-constraint handling, execution verification, and runtime overhead.

\begin{figure}[!t]
	\centering
	\subfloat[Measurement schematic.]{
		\includegraphics[width=0.48\columnwidth]{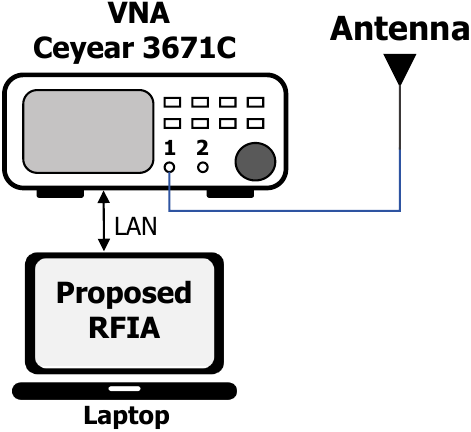}
		\label{fig:h1_antenna_schematic}
	}
	\subfloat[Prototype photograph.]{
		\includegraphics[width=0.48\columnwidth]{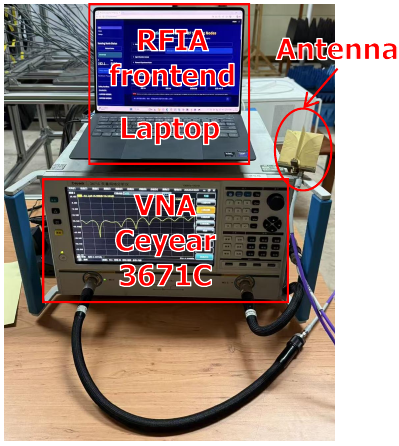}
		\label{fig:h1_antenna_photo}
	}
	\caption{Physical measurement scenario for H1. Port~1 of the VNA is connected to an antenna under test, and RFIA performs feedback-driven \(S_{11}\) scanning and resonance-frequency refinement.}
	\label{fig:h1_antenna_scenario}
\end{figure}

\begin{figure}[!t]
	\centering
	\subfloat[Measurement schematic.]{
		\includegraphics[width=0.48\columnwidth]{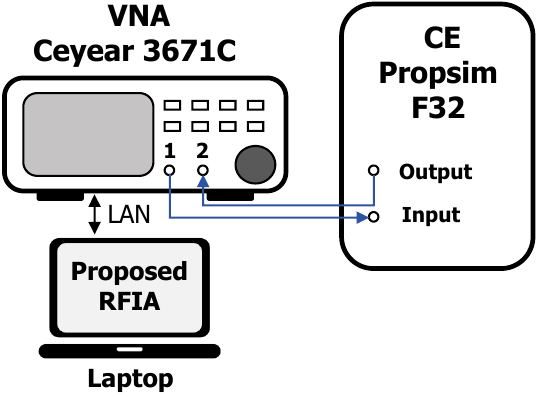}
		\label{fig:h2_channel_schematic}
	}
	\subfloat[Prototype photograph.]{
		\includegraphics[width=0.48\columnwidth]{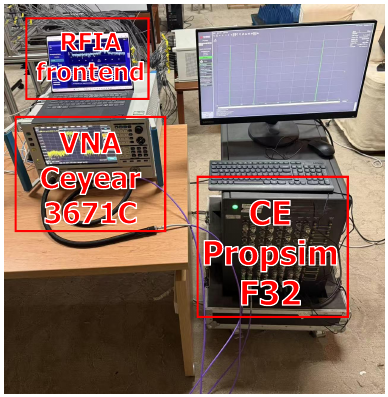}
		\label{fig:h2_channel_photo}
	}
	\caption{Physical measurement scenario for H2. The two VNA ports are connected through a channel emulator (CE), and RFIA performs channel-response acquisition and parameter estimation from the measured response.}
	\label{fig:h2_channel_scenario}
\end{figure}

\subsection{Intent Benchmark}

An intent benchmark is designed to evaluate whether the RFIA prototype can reliably handle natural-language RF-instrument tasks with different levels of complexity. As summarized in Table~\ref{tab:intent_benchmark}, the benchmark contains three difficulty levels. The easy tasks focus on basic configuration, state query, and routine \(S\)-parameter acquisition, which mainly examine the correctness of intent grounding, skill selection, and deterministic execution. The moderate tasks further require multi-step measurement, local RF analysis, result abstraction, and safety-sensitive operation handling. The hard tasks are designed to evaluate task-level RF measurement capabilities, including adaptive refinement, channel-parameter estimation, and multi-segment acquisition with data logging.

The hard tasks are associated with two representative physical measurement scenarios. In H1, port~1 of the VNA is connected to an antenna under test, as shown in Fig.~\ref{fig:h1_antenna_scenario}. RFIA is required to perform an initial wideband \(S_{11}\) scan, identify a candidate resonance interval, and adaptively refine the scan to determine the resonance frequency. This task evaluates whether the hybrid execution graph can support feedback-driven antenna measurement. In H2, the two VNA ports are connected through a CE, as shown in Fig.~\ref{fig:h2_channel_scenario}. RFIA measures the channel response around 2.5~GHz with a 40~MHz bandwidth and estimates the channel parameters from the measured response, thereby evaluating the integration of VNA acquisition, local RF analysis tools, and semantic reporting. H3 evaluates multi-band acquisition and data logging, where different frequency bands are measured with different sweep resolutions and the key measurement information is stored and reported.
For M2 and H3, the setup in Fig.~\ref{fig:h1_antenna_scenario} is extended by connecting an additional antenna to port~2, with the two antennas placed face to face to form a simple over-the-air transmission link.

Each intent is executed using the hardware-in-the-loop prototype, and the execution process is evaluated from the recorded logs. A task is regarded as successful only when the generated task contract is valid, the planning route is appropriate, the required skills, templates, or tools are available, the deterministic runtime completes all required operations without unhandled errors, and the final response satisfies the user request. For instrument-facing tasks, successful execution must be supported by readback verification, instrument-error checking, output-format validation, or tool-output validation. Therefore, the benchmark evaluates not only language understanding, but also verified task execution and measurement-level reliability.

\begin{table*}[t]
	\centering
	\caption{Intent benchmark used for prototype evaluation.}
	\label{tab:intent_benchmark}
	\renewcommand{\arraystretch}{1.15}
	\setlength{\tabcolsep}{5pt}
	\begin{tabularx}{\textwidth}{c c X}
		\toprule
		\textbf{ID} & \textbf{Difficulty} & \textbf{Intent} \\
		\midrule
		
		E1 & Easy & Set the center frequency to 3~GHz. \\
		E2 & Easy & Change the span bandwidth to 100~MHz. \\
		E3 & Easy & Set the number of sweep points to 501. \\
		E4 & Easy & Query the current number of sweep points. \\
		E5 & Easy & Query the current IF bandwidth. \\
		E6 & Easy & Query the current output power. \\
		E7 & Easy & Measure \(S_{11}\) from 3~GHz to 5~GHz. \\
		E8 & Easy & Measure \(S_{21}\) from 4~GHz to 6~GHz. \\
		E9 & Easy & Delete the local calibration set to reset the instrument. \\
		\midrule
		
		M1 & Moderate & Measure \(S_{11}\) from 3~GHz to 5~GHz and summarize the magnitude range. \\
		M2 & Moderate & Measure \(S_{21}\) from 10~GHz to 12~GHz and report the dominant peak. \\
		M3 & Moderate & Measure \(S_{11}\) from 3~GHz to 5~GHz and report the minimum magnitude. \\
		M4 & Moderate & Set the output power to 25~dBm and then measure \(S_{11}\) from 3~GHz to 5~GHz. \\
		\midrule
		
		H1 & Hard & Perform an initial wideband scan, identify the candidate resonance interval of the antenna connected to port~1, and adaptively refine the scan to determine the resonance frequency. \\
		
		H2 & Hard & Measure the channel response between ports~1 and~2 of the VNA with a center frequency of 2.5~GHz and a bandwidth of 40~MHz, and estimate the channel parameters from the measured response. \\
		
		H3 & Hard & Perform segmented \(S_{21}\) measurements with 101 points over 1--3~GHz, 501 points over 5--7~GHz, and 1001 points over 8--11~GHz; store the data in the database and report key information. \\
		
		\bottomrule
	\end{tabularx}
\end{table*}

\subsection{Overall Benchmark Results and Rule-Aware Execution}

The RFIA prototype was evaluated on all benchmark intents listed in Table~\ref{tab:intent_benchmark}, including easy, moderate, and hard tasks. These tasks cover high-frequency instrument control, state query, routine \(S\)-parameter measurement, RF-data analysis, semantic reporting, safety-sensitive operation handling, adaptive closed-loop measurement, channel-parameter estimation, and multi-segment acquisition. Table~\ref{tab:runtime_statistics} summarizes the task-level outcomes. All 16 intents were correctly interpreted, routed, and handled by RFIA. Among them, 15 tasks were physically executed or completed through deterministic tool-augmented workflows, while one safety-sensitive task was intentionally blocked by the registered calibration-protection rule.

 The easy tasks were mainly executed through direct skills or short linear workflows. E1--E3 each required a VNA connection and a single configuration action. E4--E6 included VNA connection, state query, and report generation. E7 and E8 required VNA connection, frequency-range configuration, and \(S\)-parameter measurement. E9 was routed to a safety-sensitive operation path: the VNA connection skill was executed, but the calibration-deletion operation was rejected by the registered calibration-protection rule. As a result, the destructive operation was not dispatched to the instrument.

The moderate tasks involved both instrument execution and local analysis. M1--M3 combined VNA connection, frequency-range configuration, \(S\)-parameter measurement, RF-data analysis, and summary generation. M4 tested rule-aware planning for output-power control. Although the user requested 25~dBm, the structured rule set defined a maximum allowed power of 10~dBm. The planning layer therefore modified the executable task structure by limiting the power-setting skill to 10~dBm before measurement execution. This demonstrates that safety rules are not only detected after dispatch, but can also constrain the planning result before physical operation.

The hard tasks exercised the complex-task mechanisms of RFIA. H1 was routed to a hybrid execution graph for closed-loop resonance search. Its logical execution flow is specified by the compact graph in Table~\ref{tab:resonance_execution_steps}. The runtime logger recorded 47 node evaluations for this task because the same acquisition, analysis, and refinement nodes were repeatedly evaluated during the coarse-to-fine loop; these are audit events rather than 47 manually authored task steps. H2 was routed to a tool-augmented channel-estimation workflow with local SIC analysis. H3 was routed to a multi-segment acquisition workflow, where three frequency bands were measured using different sweep resolutions before data logging and semantic reporting. These results show that RFIA can scale from short direct-skill execution to graph-based and tool-augmented RF measurement procedures.

\begin{table*}[t]
	\centering
	\caption{Runtime execution statistics for the full intent benchmark.}
	\label{tab:runtime_statistics}
	\footnotesize
	\renewcommand{\arraystretch}{1.13}
	\setlength{\tabcolsep}{3.5pt}
	\begin{tabularx}{\textwidth}{c c c c X c}
		\toprule
		\textbf{ID} & \textbf{Route} & \textbf{Logical Units} & \textbf{Log Evidence} & \textbf{Runtime Behavior} & \textbf{Outcome} \\
		\midrule
		E1 & Direct skill & 2 & 2 events & Connect VNA; set center frequency. & Completed \\
		E2 & Direct skill & 2 & 2 events & Connect VNA; set span bandwidth. & Completed \\
		E3 & Direct skill & 2 & 2 events & Connect VNA; set sweep points. & Completed \\
		E4 & Direct query & 3 & 3 events & Connect VNA; query sweep points; report result. & Completed \\
		E5 & Direct query & 3 & 3 events & Connect VNA; query IF bandwidth; report result. & Completed \\
		E6 & Direct query & 3 & 3 events & Connect VNA; query output power; report result. & Completed \\
		E7 & Linear workflow & 3 & 7 raw steps & Connect VNA; configure 3--5~GHz range; measure \(S_{11}\). & Completed \\
		E8 & Linear workflow & 3 & 7 raw steps & Connect VNA; configure 4--6~GHz range; measure \(S_{21}\). & Completed \\
		E9 & Rule-blocked path & 2 & 2 events & Connect VNA; calibration-deletion tool rejected by protection rule. & Blocked as expected \\
		\midrule
		M1 & Tool-augmented workflow & 5 & 9 raw steps & Connect VNA; configure 3--5~GHz; measure \(S_{11}\); analyze CFR; summarize. & Completed \\
		M2 & Tool-augmented workflow & 5 & 8 raw steps & Connect VNA; configure 10--12~GHz; measure \(S_{21}\); detect peak; summarize. & Completed \\
		M3 & Tool-augmented workflow & 5 & 9 raw steps & Connect VNA; configure 3--5~GHz; measure \(S_{11}\); analyze CFR; summarize. & Completed \\
		M4 & Rule-aware workflow & 5 & 11 raw steps & Requested 25~dBm was limited to 10~dBm by rule; measurement and analysis then executed. & Completed \\
		\midrule
		H1 & Hybrid execution graph & Iterative graph & 47 loop evaluations & Compact graph in Table~\ref{tab:resonance_execution_steps}; repeated coarse-to-fine \(S_{11}\) refinement. & Completed \\
		H2 & Tool-augmented workflow & 9 & 9 raw steps & Configure 2.5~GHz/40~MHz \(S_{21}\) sweep; estimate multipath parameters. & Completed \\
		H3 & Multi-segment workflow & 12 & 12 raw steps & Perform three segmented measurements with different sweep points; log data and report key information. & Completed \\
		\bottomrule
	\end{tabularx}
\end{table*}

Table~\ref{tab:rfia_overhead_summary} reports the recorded LLM token usage and runtime overhead. Each task invokes the cloud LLM once for intent understanding and high-level task abstraction, after which the executable structure is handled by the local deterministic runtime. Both MiniMax-M2.7 and Qwen3.6-27B are evaluated in hardware-in-the-loop RFIA runs. The step-signature comparison between the two runs is retained only as an audit check for generated task structure, while task success is determined by verified execution or expected rule-based rejection. The results show that the local deterministic execution remains lightweight, while latency is mainly determined by the selected cloud API and network conditions.

\begin{table}[t]
	\centering
	\caption{Recorded RFIA token usage and runtime overhead for two LLM backends.}
	\label{tab:rfia_overhead_summary}
	\footnotesize
	\renewcommand{\arraystretch}{1.15}
	\setlength{\tabcolsep}{3pt}
	\begin{tabularx}{\columnwidth}{c c c c c}
		\toprule
		\textbf{Model} & \textbf{Group} & \textbf{Success} & \textbf{Avg. Tokens} & \textbf{Planning (s)} \\
		\midrule
		MiniMax & Easy & 9/9 & 1626 & 8.13 \\
		MiniMax & Moderate & 4/4 & 2279 & 24.50 \\
		MiniMax & Hard & 3/3 & 2740 & 17.78 \\
		MiniMax & Overall & 16/16 & 1998 & 14.01 \\
		\midrule
		Qwen-27B & Easy & 9/9 & 2884 & 8.08 \\
		Qwen-27B & Moderate & 4/4 & 3040 & 14.27 \\
		Qwen-27B & Hard & 3/3 & 3131 & 14.39 \\
		Qwen-27B & Overall & 16/16 & 2969 & 10.81 \\
		\bottomrule
	\end{tabularx}
\end{table}

\subsection{Case Study I: Closed-Loop Resonance-Frequency Search}

The first hard task evaluates RFIA's feedback-control capability in antenna measurement. In this experiment, port~1 of the VNA is connected to an antenna under test, and the user requests resonance-frequency detection over 2--4~GHz. Since the final measurement band depends on the observed \(S_{11}\) response, this task cannot be represented as a fixed command sequence and is routed to a hybrid execution graph.

RFIA follows a coarse-to-fine search strategy. It first performs a wideband \(S_{11}\) scan, invokes a local RF analysis tool to detect the resonance minimum, and then iteratively updates the VNA center frequency and span until the refinement condition is satisfied. The execution flow generated by RFIA is summarized in Table~\ref{tab:resonance_execution_steps}. After several refinement iterations, RFIA converges to a resonance frequency of 3.575946~GHz with an \(S_{11}\) magnitude of \(-88.59\)~dB and a final suggested span of 1~MHz. This case shows that RFIA can use measurement feedback to refine instrument settings and produce a task-level RF result, rather than merely executing a predefined sweep.

\begin{table}[t]
	\centering
	\caption{Execution flow generated by RFIA for closed-loop resonance-frequency search.}
	\label{tab:resonance_execution_steps}
	\footnotesize
	\renewcommand{\arraystretch}{1.10}
	\setlength{\tabcolsep}{3pt}
	\begin{tabularx}{\columnwidth}{c p{1.8cm} X}
		\toprule
		\textbf{Step} & \textbf{Node Type} & \textbf{Operation} \\
		\midrule
		1 & Skill & Connect to the VNA via TCP/IP. \\
		2 & Skill & Set the center frequency to 3.0~GHz. \\
		3 & Skill & Set the sweep span to 2.0~GHz for the 2--4~GHz search range. \\
		4 & Skill & Set the number of sweep points to 1601. \\
		5 & Skill & Create and bind an \(S_{11}\) measurement. \\
		6 & Skill & Trigger a single sweep and wait for completion. \\
		7 & Skill & Read the complex \(S_{11}\) trace. \\
		8 & Skill & Read the frequency axis. \\
		9 & Tool & Detect the resonance minimum from the measured trace. \\
		10 & Condition & Check whether the depth and span refinement criteria are satisfied. \\
		11 & Skill & Refine the center frequency using the detected resonance. \\
		12 & Skill & Reduce the sweep span for the next refinement. \\
		13 & Loop & Repeat steps 6--12 until convergence or the iteration limit is reached. \\
		14 & Skill & Read the final \(S_{11}\) trace. \\
		15 & Skill & Read the final frequency axis. \\
		\bottomrule
	\end{tabularx}
\end{table}

\subsection{Case Study II: Channel-Emulator Multipath Estimation}

\begin{figure}[t]
	\centering
	\includegraphics[width=\columnwidth]{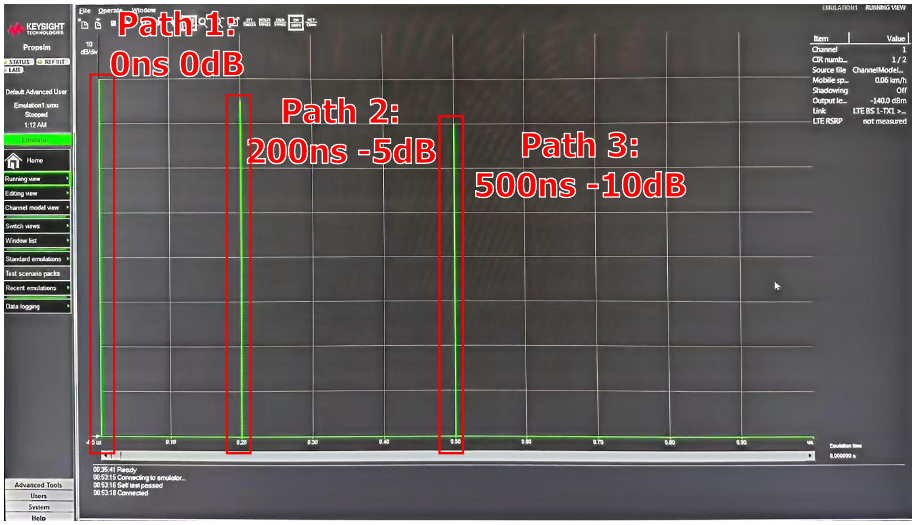}
	\caption{Configured channel impulse response loaded in the CE.}
	\label{fig:CE_setting_photo}
\end{figure}

The second challenging task evaluates the capability of RFIA to integrate VNA data acquisition with local RF-domain analysis. In this experiment, the two ports of the VNA are connected via a CE. The CE is configured to generate three dominant propagation paths with time delays of 0 ns, 200 ns, and 500 ns and corresponding power levels of 0 dB, -5 dB, and -10 dB, as depicted in Fig. \ref{fig:CE_setting_photo}. The center frequency is set to 2.5 GHz and the bandwidth to 40 MHz. 

The user asks RFIA to measure the channel response with a center frequency of 2.5~GHz, a bandwidth of 40~MHz, and 1001 sweep points, and to estimate the multipath parameters from the measured response. RFIA classifies the task as a measurement-analysis task and routes it to a tool-augmented workflow. The execution structure consists of nine main steps: VNA connection, center-frequency configuration, span configuration, sweep-point configuration, \(S_{21}\) measurement creation, single-sweep triggering, complex trace acquisition, frequency-axis acquisition, and local SIC-based multipath estimation. The corresponding execution flow is summarized in Table~\ref{tab:h2_execution_steps}. All instrument-facing operations are dispatched through verified skills and checked before the next step is executed.

\begin{table}[t]
	\centering
	\caption{Execution flow of the H2 channel-parameter estimation task.}
	\label{tab:h2_execution_steps}
	\footnotesize
	\renewcommand{\arraystretch}{1.15}
	\setlength{\tabcolsep}{3pt}
	\begin{tabularx}{\columnwidth}{c p{2.2cm} X}
		\toprule
		\textbf{Step} & \textbf{Node Type} & \textbf{Operation} \\
		\midrule
		1 & Skill & Connect to the VNA through the LAN-based SCPI interface. \\
		2 & Skill & Set the center frequency to 2.5~GHz. \\
		3 & Skill & Set the sweep span to 40~MHz. \\
		4 & Skill & Set the sweep points to 1001. \\
		5 & Skill & Create and bind an \(S_{21}\) measurement. \\
		6 & Skill & Trigger a single sweep and wait for completion. \\
		7 & Skill & Read complex \(S_{21}\) data. \\
		8 & Skill & Read the frequency axis. \\
		9 & Tool & Estimate multipath parameters using the local SIC tool. \\
		\bottomrule
	\end{tabularx}
\end{table}

The local SIC tool first estimates and removes the bulk fixed delay introduced by the measurement chain and CE \cite{wang2025multi}. It then estimates the relative delays and powers of the dominant multipath components from the measured frequency response. The estimated fixed delay is 4504.145~ns. After bulk-delay alignment, RFIA identifies three dominant paths, which agrees with the three-path configuration of the CE. The estimated relative delays and powers are shown in Table~\ref{tab:h2_multipath_result}.

\begin{table}[t]
	\centering
	\caption{Estimated multipath parameters for the H2 channel-emulator task.}
	\label{tab:h2_multipath_result}
	\footnotesize
	\renewcommand{\arraystretch}{1.15}
	\setlength{\tabcolsep}{4pt}
	\begin{tabularx}{\columnwidth}{c c c c}
		\toprule
		\textbf{Path} & \textbf{Rel. Delay (ns)} & \textbf{Abs. Delay (ns)} & \textbf{Rel. Power (dB)} \\
		\midrule
		1 & 0.000 & 4504.145 & 0.00 \\
		2 & 200.100 & 4704.245 & \(-5.18\) \\
		3 & 500.350 & 5004.495 & \(-10.35\) \\
		\bottomrule
	\end{tabularx}
\end{table}

The fitting residual is \(-21.58\)~dB and the explained energy ratio is 0.993, indicating that the estimated three-path model explains most of the measured response. The fit is reported as reliable by the local tool. The final user-facing answer generated by RFIA is based on the structured tool output and reports the number of paths, the estimated fixed delay, the relative delays, the relative powers, and the fitting residual. This case demonstrates that RFIA can perform RF-domain parameter estimation through local deterministic tools while using the LLM only for intent understanding, orchestration, and natural-language summarization.

\subsection{Case Study III: Multi-Segment Measurement and Data Logging}

The third hard task evaluates RFIA's ability to organize heterogeneous measurement configurations within a single user intent. The task requires segmented \(S_{21}\) measurements over 1--3~GHz, 5--7~GHz, and 8--11~GHz with 101, 501, and 1001 sweep points, respectively, followed by database storage and key-information reporting. Unlike H1, this task is not closed-loop, but it requires the planner to coordinate multiple frequency segments, sweep resolutions, data acquisition steps, and logging operations in a consistent workflow.

RFIA generates a 12-step runtime workflow consisting of 10 skill nodes and 2 tool nodes. The skill nodes connect to the VNA, configure the three measurement segments, execute the corresponding sweeps, and acquire the measured data, while the tool nodes store the acquired traces and generate a compact segment-level report. The successful completion of H3 demonstrates that RFIA can bind different sweep settings within one task, preserve the dependencies among configuration, acquisition, data packaging, and logging, and return structured measurement information.

\subsection{LLM Baseline and Model-Size Comparison}

To isolate the architectural contribution of RFIA, a direct LLM-to-SCPI baseline was evaluated using MiniMax-M2.7 and the same 16 benchmark intents. The model was prompted to generate SCPI command sequences for a Ceyear 3671C VNA controlled through a LAN SCPI socket, without access to RFIA's verified skills, workflow templates, state manager, safety rules, or local RF-analysis tools. The generated commands were archived and evaluated offline according to task-specific criteria, without being dispatched to the physical instrument.

\begin{figure*}[t]
	\centering
	\includegraphics[width=0.90\textwidth]{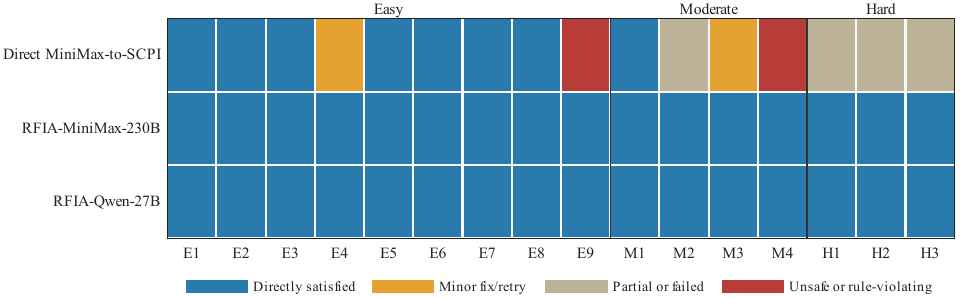}
	\caption{Task-level outcome map over the 16 benchmark intents. Direct MiniMax-to-SCPI code generation often requires manual intervention and fails on rule-protected, feedback-dependent, or RF-analysis-intensive tasks, whereas RFIA succeeds with both MiniMax-M2.7 and Qwen3.6-27B in hardware-in-the-loop execution.}
	\label{fig:experiment_outcome_map}
\end{figure*}

The comparison examines whether LLM-generated code can directly support reliable RF measurement automation, rather than only simple command translation. As shown in Fig.~\ref{fig:experiment_outcome_map}, RFIA was evaluated with both MiniMax-M2.7 and the smaller SiliconFlow-hosted Qwen3.6-27B model. Both RFIA variants completed all 16 benchmark tasks in the hardware-in-the-loop setup under the same verification and rule policies. Their step-level signatures were also compared as a structural audit, confirming that the smaller model produced executable plans consistent with the MiniMax-M2.7 RFIA reference.

In the direct baseline, only eight tasks were directly usable without manual intervention. Two additional tasks required manual correction or repeated attempts, namely E4 and M3. The remaining failures fall into three categories. First, E9 and M4 violated safety or protection requirements because the baseline did not enforce calibration-deletion and output-power rules. Second, M2 and H2 relied on unverified LLM interpretation instead of deterministic RF-analysis tools, leading to unreliable or incorrect RF-domain results. Third, H1 and H3 were not correctly completed because direct generation did not preserve feedback validation or segment-level measurement timing.

These results show that RFIA's advantage is not merely improved SCPI-generation accuracy, but a categorical improvement in safety enforcement, deterministic RF-analysis integration, and feedback-dependent execution. Direct LLM-to-SCPI generation can draft simple command sequences, but it cannot provide the execution guarantees required for reliable RF measurement automation. By confining the LLM to intent understanding and high-level planning, RFIA allows both large and smaller models to drive the same verified runtime, while all instrument-facing operations remain under deterministic execution.

\subsection{Discussion}
The experimental results highlight the complementary roles of the LLM and the deterministic runtime. The LLM provides task understanding, abstraction, and summarization, while verified skills and local tools handle instrument operations and RF-domain computations. This separation enables natural-language usability without exposing the physical instrument to unconstrained model outputs. Since the current prototype uses cloud LLM APIs, the recorded planning latency is affected by network conditions. The compatibility with smaller models like Qwen3.6-27B observed in our hardware-in-the-loop tests suggests that future on-premise deployments could use compact, fine-tuned models to reduce cloud dependency and latency.

\section{Conclusion}
This paper presented RFIA, a lightweight natural-language agent framework for reliable task-driven RF instrument control. RFIA separates LLM-based task understanding and high-level planning from deterministic instrument execution. By organizing verified skills, workflow templates, RF analysis tools, instrument-specific rules, and retrieval-assisted SCPI knowledge into a structured knowledge base, RFIA constrains executable operations before any instrument-facing command is dispatched, providing a practical alternative to direct LLM-to-SCPI generation.
A hardware-in-the-loop prototype was implemented on a commercial VNA and evaluated using a 16-intent benchmark covering configuration, query, \(S\)-parameter acquisition, rule-aware operation, RF-data analysis, adaptive resonance search, channel-parameter estimation, and multi-segment acquisition. RFIA handled all benchmark intents under predefined execution and safety policies with both MiniMax-M2.7 and Qwen3.6-27B, including one expected safety rejection. These results indicate that the proposed separation between LLM-based task abstraction and verified runtime execution can support reliable RF measurement automation across different cloud LLM backends, while retaining compatibility with existing SCPI-enabled instruments.

	\bibliographystyle{IEEEtran}
	\bibliography{TM}
	\addcontentsline{toc}{section}{\refname}
\end{document}